\documentclass[conference]{IEEEtran}

\usepackage{cite}
\usepackage{dcolumn}
\usepackage{insertfig}
\usepackage{revhistory}
\usepackage{times}
\usepackage{url}
\usepackage{bm}
\usepackage{amsmath}

\newcommand{\dd}{\mathrm{d}}
\newcommand{\diag}[1]{{\rm diag}(#1)}

\makeindex             

\renewcommand{\insertfigsize}{0.37}
\renewcommand{\insertfigssize}{0.74}

\begin{document}

 \title{Information Propagation Analysis of Social Network Using the Universality of Random Matrix}


\author{\authorblockN{Yusuke Sakumoto\authorrefmark{1},
Tsukasa Kameyama\authorrefmark{2},
Chisa Takano\authorrefmark{3} and
Masaki Aida\authorrefmark{1}}
\authorblockA{\authorrefmark{1}Tokyo Metropolitan University, 6-6 Asahigaoka, Hino-shi, Tokyo, Japan 191-0065\\ Email: \{sakumoto,maida\}@tmu.ac.jp}
\authorblockA{\authorrefmark{2}Nochu Information System Co., Ltd, 3-2 Toyosu, Koto, Tokyo 135-0061, Japan\\
Email: kameyama445@gmail.com}
\authorblockA{\authorrefmark{3}Hiroshima City University, 3-4-1 Ozukahigashi, Asaminami, Hiroshima 731-3166\\
Email: takano@hiroshima-cu.ac.jp}}

\maketitle

\begin{abstract}
Spectral graph theory gives an algebraical approach to analyze the
dynamics of a network by using the matrix that represents the network
structure.  However, it is not easy for social networks to apply the
spectral graph theory because the matrix elements cannot be given
exactly to represent the structure of a social network.  The matrix
element should be set on the basis of the relationship between
persons, but the relationship cannot be quantified accurately from
obtainable data~(e.g., call history and chat history).  To get around
this problem, we utilize the universality of random matrix with the
feature of social networks.  As such random matrix, we use normalized
Laplacian matrix for a network where link weights are randomly given.
In this paper, we first clarify that the universality~(i.e., the
Wigner semicircle law) of the normalized Laplacian matrix appears in
the eigenvalue frequency distribution regardless of the link weight
distribution.  Then, we analyze the information propagation speed by
using the spectral graph theory and the universality of the normalized
Laplacian matrix.  As the results, we show that the worst-case speed
of the information propagation changes at most 2 if the
structure~(i.e., relationship among people) of a social network
changes.
\end{abstract}


\section{Introduction}

The emergence of social networking services~(SNSs) and the widespread
of mobile devices promote people interaction beyond anticipation in
the society.  As the results, the people interaction has strong
capability to propagate the information submitted by someone to the
whole society.  In the recent years, it is important for the success
of a new product and a new spot to propagate its information all over
the social network through not only face-to-face offline conversation
but also online communication via SNS~(e.g., Twitter and Instagram).
Therefore, the understanding of the information propagation property
on social networks is essential to design marketing strategy of new
products and new spots.

Spectral graph theory~\cite{Chun} is widely used to analyze the
property of network dynamics by using the eigenvalues and the
eigenvectors of a matrix~(e.g., Laplacian matrix) that represents the
structure of the network.  However, when applying the spectral graph
theory to social network analysis, there is the difficulty due to two
following reasons.  First, social networks are huge.  To analyze them
using spectral graph theory, the eigenvalues and eigenvectors of the
huge-size matrix must be calculated but this calculation is impossible
because of its computational cost.  Secondly, the relationship between
persons in a social network is complex.  It is hard to quantify the
relationship accurately from obtainable data (e.g., call history and
chat history).  To represent the social network structure by a matrix,
the matrix elements must be given exactly on the basis of the
relationship but we need to overcome the difficult task of the
relationship quantification.  Therefore, before applying the spectral
graph theory to the social network analysis, we should discuss the way
around the above-mentioned problem.

A random matrix is a matrix whose elements are random variables, and
has been utilized to analyze large-scale and complex structure in
quantum mechanics~\cite{random_matrix1,Wigner}.  In quantum mechanics,
there is a method to derive the electron orbital around an atomic
nucleus by using a matrix that represents the atom structure.
However, for a large atom~(e.g., uranium) having many electrons with
complex orbital, it is impossible to give the matrix elements exactly.
Hence, quantum mechanics gives up representing such large and complex
atom structure exactly, and analyze electron orbital property using
the universality when the matrix elements are given by random
variables.  The analysis using the universality of the random matrix
has had great success in quantum mechanics.  The circumstance of the
large and complex atom analysis in quantum mechanics is like the
social network analysis, so random matrix would solve the fundamental
problem in the social network analysis.

In this paper, we first investigate the universality of random matrix
with the feature of social networks.  As such random matrix, we use
normalized Laplacian matrix for a network where link weights are
randomly given.  We clarify that the universality~(i.e., the Wigner
semicircle law) of the normalized Laplacian matrix appears in the
eigenvalue frequency distribution regardless of the link weight
distribution in random networks generated with the popular
models~(i.e., ER~(Erd{\"o}s--R{\'e}nyi) model~\cite{ER} and
BA~({Barab\'asi--Albert})~\cite{BA} model), which are also used in
several studies~\cite{ER, BA, degree, Jayen, Fark, Sakumoto}.  Then,
we analyze the information propagation speed in social networks by
using the clarified universality.  In this analysis, we model the
information propagation by a random walk in the light of the
resemblance between their characteristics.  Since random walks are too
slower than the information propagation in a social network, our
analysis focuses on the worst-case situation.  As a metric for the
information propagation speed, we use the expected value of first
arrival times of the random walker for each node.  As the result of
our analysis using spectral graph theory and the clarified
universality, we show that the worst-case speed of the information
propagation changes at most 2 if the structure~(i.e., relationship
among people) of a social network changes.

This paper is organized as follows.  In Sect.~\ref{sec:pre}, we
describe the normalized Laplacian matrix and the Wigner semicircle law
as the preliminary of our discussion.  In Sect.~\ref{sec:val}, we
generate the random matrix with the social network feature, and
investigate its universality.  Section~\ref{sec:ana} analyzes the
information propagation property using the universality clarified in
Sect.~\ref{sec:val}.  Finally, in Sect.~\ref{sec:conclusion}, we
conclude this paper and discuss future works.

\section{Preliminary}
\label{sec:pre}


\subsection{Normalized Laplacian Matrix}

We denote an undirected network with $n$ nodes by $G(V, E)$ where $V$
and $E$ are the sets of nodes and links, respectively.
Let $\bm{A} = (A_{ij})_{1 \le i,j \le n}$ be the adjacency matrix, which represents the
link structure of network $G$.  $A_{ij}$ is defined by 
\begin{align}
   A_{ij} := 
               \begin{cases}
                  w_{ij} \quad  &( (i,j) \in E ) \\
                  0        &( (i,j) \notin E )
               \end{cases},
               \label{eq:adjacency matrix}
\end{align}
where $w_{ij} > 0$ is the weight of link $(i, j)$.  Since network $G$
is undirected, adjacency matrix $\bm{A}$ is symmetric~$(A_{ij} = A_{ji})$.
Let $\bm{D}= \diag{d_{i}}_{1 \le i \le n}$ be the degree matrix where
$d_i = \sum_{j = 1}^n w_{ij}$ is the weighted degree of node $i$.
Using adjacency matrix $\bm{A}$ and degree matrix $\bm{D}$, Laplacian
matrix $\bm{L}$ for network $G$ is defined by
\begin{align}
  \bm{L} := \bm{D}-\bm{A}.
  \label{eq:Laplacian matrix}
\end{align}
Laplacian matrix $\bm{L}$ represents the node and link structure of
network $G$.


Normalized Laplacian matrix $\bm{N}$ is also used to represent the
network structure.  $\bm{N}$ is defined by
\begin{align}
  \bm{N} := \bm{D}^{-1/2} \,\bm{L} \,\bm{D}^{-1/2}.
  \label{eq:N}
\end{align}
Since $\bm{N}$ is symmetric, $\bm{N}$ can  be always diagonalized.
Hence, $\bm{N}$ is also given by
\begin{align}
  \bm{N} = \bm{P}\,\bm{\Lambda}\,\bm{P}^{-1},
  \label{eq:Laplacian}
\end{align}
where $\bm{\Lambda} = \diag{\lambda_l}_{1 \le l \le n}$ and $\bm{P} =
(\bm{q}_l)_{1 \le l \le n}$.  $\lambda_l$ and $\bm{q}_l$ are $l$-th
eigenvalue of $\bm{N}$ and the eigenvector for $\lambda_l$,
respectively.  In this paper, we arrange eigenvalue $\lambda_l$ in
ascending order~(i.e., $0 = \lambda_1 < \lambda_2 \le ... \le
\lambda_n < 2$).  Since $\bm{N}$ is symmetric, eigenvector
$\bm{q}_l$ is the orthonormal basis.  Namely, $\sum_{i = 1}^n
q_k(i)\,q_l(i) = \delta_{kl}$ where $\delta_{kl}$ is the Kronecker
delta.  According to Eq.~(\ref{eq:Laplacian}), the combination of
$\bm{P}$ and $\bm{\Lambda}$ also represent the network structure
equivalent to $\bm{N}$.

In Sect.~\ref{sec:ana}, we discuss the relation between the random
walk and the information propagation in a social network, and analyze
the information propagation based on $\bm{N}$.

%
%
%


\subsection{Wigner Semicircle Law}

The Wigner semicircle law~\cite{Wigner} is the universality that
appears in the eigenvalue density distribution of random matrices.
Let $\bm{X} = (X_{ij})_{i \le i,j \le n}$ be a $n \times n$ real
symmetric matrix where $X_{ij}$ is a random variable.  $X_{ij}$ for $j
\geq i$ follows an independent identical distribution where all
odd-order moments are zero and all even-order moments are finite
amounts.
We denote $k$-th eigenvalue for a sample of $\bm{X}$ by
$\lambda_k$~$(k=1,\,\dots,\,n)$.
We look at the eigenvalue density $\rho_n(\lambda)$, which is given by
\begin{align}
   \rho_n(\lambda) = \frac{1}{n}\,\sum_{k=1}^{n} \delta \biggl(\lambda-\frac{\lambda_k}{\sqrt{n}} \biggr),
   \label{eq:randommatrix_ev}
\end{align}
where $\delta(x)$ is the Dirac delta function of $x$.
As the limit of $\rho_n(\lambda)$ with $n \rightarrow \infty$,
$\rho_X(\lambda)$ follows by
\begin{align}
  \rho_X(\lambda) &= \lim_{n \rightarrow \infty} \rho_n(\lambda) \nonumber\\
                  &= \begin{cases}
                  \displaystyle \frac{1}{2\,\pi\,\sigma^2} \sqrt{4 \sigma^2 - \lambda^2} & (|\lambda|<2\sqrt{\sigma^2}) \\  
                 0 & (\mathrm{otherwise})   
               \end{cases},
               \label{eq:Wigner}
\end{align}
where $\sigma$ is the standard deviation of the distribution of
$X_{ij}$.  Because of $\int_{-\infty}^{\infty} \rho_X(\lambda) \, \dd
\lambda = 1$, $\rho_X(\lambda)$ is the density distribution of the
eigenvalues.  The Wigner semicircle law means that the eigenvalue
density distribution is given by Eq.~(\ref{eq:Wigner}).


\section{Random Matrix with Social Network Feature}
\label{sec:val}

The existing study~\cite{existing} clarified the universality of the
well-known networks~(ER network~\cite{ER} and BA network~\cite{BA})
with unweighted links.  The clarified universality says that the
eigenvalue frequency distribution of normalized Laplacian matrix
$\bm{N}$ for an unweighted ER and BA network satisfies the Wigner
semicircle law if the network fulfills the degree condition
$k^{2}_\mathrm{min} \gg k_\mathrm{ave}$ where $k_\mathrm{min}$ and
$k_\mathrm{ave}$ are the minimum and average of degrees~(numbers of
links from a node), respectively.

To represent the structure of a social network, links should be
weighted on the basis of the relationship between persons since people
have diverse relationships in the social network.  However, the
relationships are complex, so we cannot give the link weights exactly.
Hence, the universality for weighted networks should be needed for
social network analysis.

In this section, we clarify the universality for weighted random
networks to analyze the social network property.  We first generate
normalized Laplacian matrix $\bm{N}$ for randomly-weighted random
networks~(ER network and BA network) as random matrix, and then
investigate the universality of $\bm{N}$.

\subsection{Generation Method of Random Matrix}

As a random matrix with the feature of social networks, we use
normalized Laplacian matrix $\bm{N}$ for randomly-weighted random
networks~(ER network and BA network) generated by the following steps.
We denote the existing probability of links in the ER network by $p$.
Then, let $n_0$ and $n_a$ be the number of initial connected nodes and
the number of adding links from a new node in the BA network,
respectively.

\begin{enumerate}
    \item Input expected average number $k_{\rm ave}$ of links of each
      node.
    \item Generate an unweighted random network~($w_{ij} =1$) based on
      the ER model or the BA model.
      \begin{enumerate}
        \item When we generate a ER network, we set existing
          probability $p$ by $k_{\rm ave}/(n-1)$ so that each node has
          $k_{\rm ave}$ links in average.
        \item When we generate a BA network, we set $n_0$ and $n_a$ so
          that the BA network has about $n\, k_{\rm ave}/(1-q)$ links.
          After generating the BA network, we cut the links randomly
          with the probability $q$.  Each node in the cut network
          has $k_{\rm ave}$ in average.  We call the cut BA
          network simply as ``BA network''.
      \end{enumerate}
    \item Generate random values that follow a probability
      distribution~(e.g., constant distribution, uniform distribution
      and exponential distribution), and set link weight $w_{ij}$ for
      $i > j$ and $(i, j) \in E$ by the random value.
\end{enumerate}

Figure~\ref{fig:dist-deg_ba_vs_er} shows the degree~(i.e., number of
links) distribution of ER and BA networks obtained from the above
steps with $n = 1,000$, $k_{\rm ave} = 20$, $q = 0.5$, and $w_{ij} =
1$ for all $(i, j) \in E$.  As reference, we drew the distribution of
the original BA network~\cite{BA} in Fig.~\ref{fig:dist-deg_ba_vs_er}.
In the ER network and the BA network, each node has 20 links in
average.  On the contrary, the average links of the original BA
network is about 40.  According to Fig.~\ref{fig:dist-deg_ba_vs_er},
the degree distribution for large degree nodes in the BA network has
the same scaling exponent~(-3) of the original BA network.  Hence,
the cutting of links in step 2b keeps the scale-free property of
original BA networks.

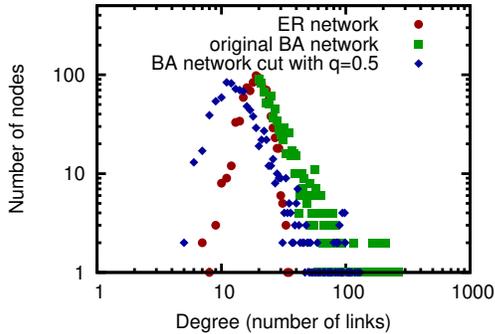
\insertfig{dist-deg_ba_vs_er}{Degree distribution of unweighted ER network and BA network}

The reason why we cut the links of BA networks randomly in step 2b is
described below.  According to the BA model~\cite{BA}, minimum degree
$k_\mathrm{min}$ and average degree $k_\mathrm{ave}$ of original BA
networks are always the same value when keeping the configuration of
$n_0$ and $n_a$.  On the contrary, $k_\mathrm{min}$ and
$k_\mathrm{ave}$ of ER networks are randomly changed when keeping the
configuration of $p$.  By cutting the links of BA networks randomly,
the BA networks have different $k_\mathrm{min}$ and $k_\mathrm{ave}$,
and we can compare the results for BA networks and ER networks under
comparable condition.

\subsection{The Universality and its Applicable Condition}

We experimentally investigate the eigenvalue frequency distribution
$f_N(\lambda)$ of normalized Laplacian matrix $\bm{N}$ for the
randomly-weighted random networks, and clarify the universality~(the
Wigner semicircle law) of $f_N(\lambda)$ and the applicable condition.

In the investigation, we use the constant distribution with $w_{ij} =
\overline{w}$, the uniform distribution with the range
$[\overline{w}/2, \,3\,\overline{w}/2]$ and exponential distribution
with the average $\overline{w}$ to set link weight $w_{ij}$ randomly.
Note that these distributions have the same average of link weights,
but a different variance of link weights.  By the comparison of the
results for the different distributions, we clarify the universality
regardless of link weights in social networks.  We repeat the
generation of $\bm{N}$ 100 times, and calculate the average of these
results.  We use the parameter configuration shown in
Tab.~\ref{tab:param} as default.

\begin{table}[t]
  \caption{Default parameter configuration}
  \begin{center}
   \scriptsize
   \begin{tabular}{|l|c|c|}
     \hline
     parameter & symbol & configuration\\
     \hline
     number of nodes & $n$ & 1,000 \\
     distribution of link weight $w_{ij}$ & & uniform distribution\\
     average of link weights & $\overline{w}$ & 1\\
     number of bins in $f_N(\lambda)$ & $n_h$ & 50 \\
     \hline
   \end{tabular}
   \label{tab:param}
 \end{center}
\end{table}

We discuss the applicable condition of the universality on the basis
of $k^{2}_\mathrm{min}$ and $k_\mathrm{ave}$ like the existing
study~\cite{existing}.  In
Fig.~\ref{fig:avg-deg_vs-min-deg_ba_vs_er_N=1000Xk=40}, we first show
$k^{2}_\mathrm{min}$ and $k_\mathrm{ave}$ in randomly-weighted ER and
BA networks with different expected average number of links, $k_{\rm
  ave}$, and the uniform distribution.  As $k_{\rm ave}$ increases,
$k^{2}_\mathrm{min}$ and $k_\mathrm{ave}$ increases simultaneously,
but the increasing speed of $k^{2}_\mathrm{min}$ is larger than that
of $k_\mathrm{ave}$.  According to
Fig.~\ref{fig:avg-deg_vs-min-deg_ba_vs_er_N=1000Xk=40}, in order to
fulfill the degree condition $k^{2}_\mathrm{min} \gg k_\mathrm{ave}$
in~\cite{existing}, $k_\mathrm{ave} \ge 16$ for ER network and
$k_\mathrm{ave} \ge 24$ for BA network are needed at least,
respectively.  In this paper, we set the average of link weight
$w_{ij}$ to an amount equal to or greater than 1.  Hence, if
$k^{2}_\mathrm{min} \gg k_\mathrm{ave}$ is fulfilled,
$d^{2}_\mathrm{min} \gg d_\mathrm{ave}$ is also fulfilled in average.
Hence, we use $k^{2}_\mathrm{min} \gg k_\mathrm{ave}$ instead of
$d^{2}_\mathrm{min} \gg d_\mathrm{ave}$ in order to clarify the
applicable condition regardless of link weights.

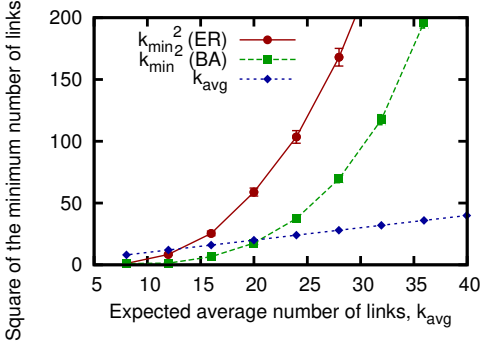
\insertfig{avg-deg_vs-min-deg_ba_vs_er_N=1000Xk=40}{The expected average number $k_\mathrm{ave}$ of links vs. the square of the minimum number $k_\mathrm{min}$ of links in randomly-weighted ER network and BA network}

In Figs.~\ref{fig:dist-eig_er_N=1000Xk=40Xs=1Xplt} and
\ref{fig:dist-eig_cut-ba_N=1000Xk=40Xs=1Xplt}, we show eigenvalue
frequency distributions $f_N(\lambda)$ of normalized Laplacian matrix
$\bm{N}$ for randomly-weighted ER and BA networks, respectively.  When
we obtained $f_N(\lambda)$, we first counted the number of eigenvalues
of $\bm{N}$ within $[\lambda - h_b/2, \lambda + h_b/2]$ where $h_b =
(\lambda_n - \lambda_2)/n_h$.  Then, we normalized the counted number
so that $\sum_{i = 1}^{n_h} f_N(\theta_i) = 1$ where $\theta_i =
\lambda_2 + (i-1/2)\,h_b$, and $n_h$ is the number of bins in
$f_N(\lambda)$.  Note that we took out the minimum eigenvalue
$\lambda_1$ of $\bm{N}$ when obtaining $f_N(\lambda)$ because
$\lambda_1$ is always 0.  In
Figs.~\ref{fig:dist-eig_er_N=1000Xk=40Xs=1Xplt} and
\ref{fig:dist-eig_cut-ba_N=1000Xk=40Xs=1Xplt}, we also draw the
semicircle distribution, which is given by
\begin{align}
   \rho_N(\lambda) =
               \begin{cases}
                 \displaystyle \frac{2}{\pi\,r^2} \sqrt{r^2-(\lambda-1)^{2}} &\!\! (\lambda_2<\lambda<\lambda_n) \\  
                 0 & \!\!(\mathrm{otherwise})   
               \end{cases},
               \label{eq:semicircle}
\end{align}
where $r$ is the radius of the semicircle distribution, and is given
by $1 - \lambda_2$ or $\lambda_n - 1$.  According to the range of
$\lambda_2$ and $\lambda_n$, $r$ must be within $0 < r < 1$.
Equation~(\ref{eq:semicircle}) is essentially equivalent to
Eq.~(\ref{eq:Wigner}) in the Wigner semicircle law because there is
just the difference of the semicircle center.  Hence, we define that
the Wigner semicircle law is satisfied if eigenvalue frequency
distribution $f_N(\lambda)$ coincides with the semicircle distribution
given by Eq.~(\ref{eq:semicircle}).  According to
Figs.~\ref{fig:dist-eig_er_N=1000Xk=40Xs=1Xplt} and
\ref{fig:dist-eig_cut-ba_N=1000Xk=40Xs=1Xplt}, eigenvalue frequency
distribution $f_N(\lambda)$ of $\bm{N}$ for the randomly-weighted ER
and BA networks with $k_{\rm ave} = 8$ differs from the semicircle
distribution (\ref{eq:semicircle}), but the distributions for the
randomly-weighted ER and BA networks with $k_{\rm ave} = 36$ almost
coincide with it.  Hence, we expect that large $k_{\rm ave}$ is needed
to satisfy the Wigner semicircle law.

\insertfigs{dist-eig_er_N=1000Xk=40Xq=0X2Xs=1Xplt}{$k_{\rm ave} = 8$}
{dist-eig_er_N=1000Xk=40Xq=0X9Xs=1Xplt}{$k_{\rm ave} = 36$}
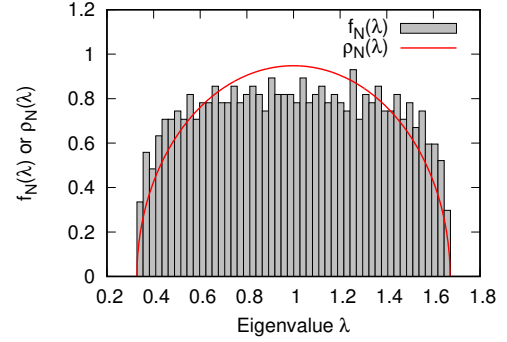
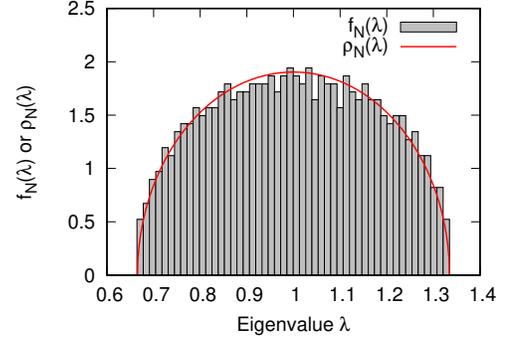
{dist-eig_er_N=1000Xk=40Xs=1Xplt}{The eigenvalue frequency distributions of {$\bm{N}$}'s for randomly-weighted ER networks}

\insertfigs{dist-eig_cut-ba_N=1000Xk=40Xq=0X2Xs=1Xplt}{$k_{\rm ave} = 8$}
{dist-eig_cut-ba_N=1000Xk=40Xq=0X9Xs=1Xplt}{$k_{\rm ave} = 36$}
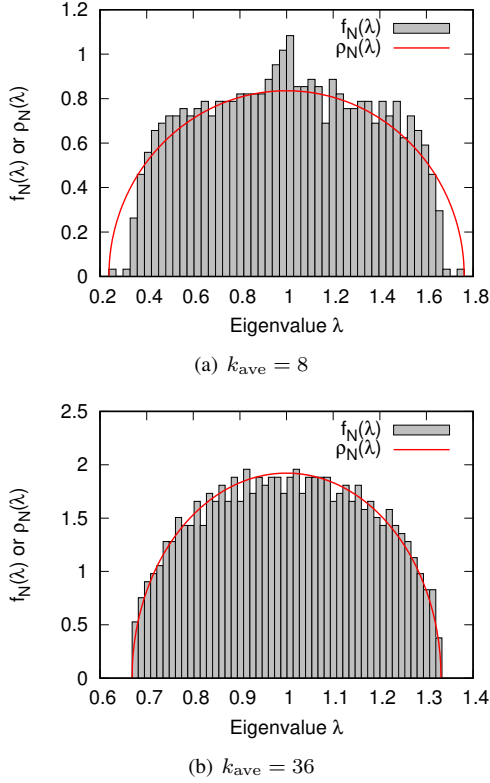
{dist-eig_cut-ba_N=1000Xk=40Xs=1Xplt}{The eigenvalue frequency distributions of {$\bm{N}$}'s for randomly-weighted BA networks}

In order to clarify whether the eigenvalue frequency distribution
$f_N(\lambda)$ satisfies the Wigner semicircle law, we investigate the
difference between $f_N(\lambda)$ and $\rho_N(\lambda)$.  As the
definition of the difference, we use relative error $\epsilon$.
Relative error $\epsilon$ is given by
\begin{align}
\epsilon = \frac{1}{n_h} \sum_{i=1}^{n_h} \frac{|f_N(\theta_i) - P(\theta_i)|}{P(\theta_i)}
   \label{eq:誤差率1},
\end{align}
where $\theta_i = \lambda_2 + (i-1/2)\,h_b$, and $P(\theta_i)$ is the
probability that the eigenvalues of $\bm{N}$ is within $[\theta_i -
  h_b/2, \theta_i + h_b/2]$.  Namely, $P(\theta_i)$ is defined by
\begin{align}
  P(\theta_i) = \int_{\theta_i - h_b/2}^{\theta_i + h_b/2} \rho_N(\lambda) \,\dd \lambda.
\end{align}
Note that $P(\theta_i)$ should become the true value of
$f_N(\theta_i)$ for $n \rightarrow \infty$ and $n_h \rightarrow
\infty$.

In Fig.~\ref{fig:err-dist-eig_ba_vs_er_N=1000Xk=40}, we show the
results of $\epsilon$ for randomly-weighted ER and BA networks with
the constant distribution, the uniform distribution, and the
exponential distribution of link weight $w_{ij}$.  As $k_{\rm ave}$
increases, $\epsilon$ decreases regardless of the link weight
distribution and network topology~(ER or BA).  The reason why
$\epsilon$ does not approach to 0 is that there are discretization
error due to the finite setting of $n$ and $n_h$.  Moreover,
$\epsilon$ for the uniform and exponential distributions approaches to
that for the constant distribution.  According to \cite{existing},
$f_N(\lambda)$ with the constant distribution~($w_{ij} = 1$) satisfies
the Wigner semicircle law when the degree condition
$k^{2}_\mathrm{min} \gg k_\mathrm{ave}$ is fulfilled.  Hence,
$f_N(\lambda)$ with the uniform and exponential distributions also
satisfies the Wigner semicircle law.

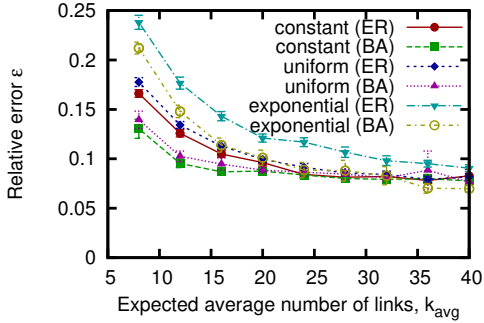
\insertfig{err-dist-eig_ba_vs_er_N=1000Xk=40}{Relative error $\epsilon$ of eigenvalue frequency distribution $f_N(\lambda)$ for randomly-weighted ER and BA networks}

From the above results, we can find that if the degree condition
$k^{2}_\mathrm{min} \gg k_\mathrm{ave}$ is fulfilled, the eigenvalue
frequency distribution of normalized Laplacian matrix $\bm{N}$ for
randomly-weighted ER and BA networks satisfies the Wigner semicircle
law given by Eq.~(\ref{eq:semicircle}), which is the semicircle
distribution determined by only the second smallest eigenvalue
$\lambda_2$ or the maximum eigenvalue $\lambda_n$ of $\bm{N}$.  Hence,
$\lambda_2$ and $\lambda_n$ are important to understand the social
network property in the a case fulfilling the degree condition.

\section{Analysis of the Information Propagation Speed}
\label{sec:ana}

In this section, we analyze the speed of the information
propagation on social networks fulfilling the degree condition
$k^{2}_\mathrm{min} \gg k_\mathrm{ave}$.  If the degree condition is
not fulfilled in a social network, there are many persons with too
small number of friends.  However, such persons would be minority in
an actual social network, and contribute less to the information
propagation on the entire social network.  Therefore, we ignore such
persons, and focus on social networks fulfilling the degree condition.

\subsection{Metric of the Information Propagation Speed}

The information propagation in a social network is involved with the
chain of word-of-mouth communications~(e.g., face-to-face offline
conversation, and online communication via SNS) between persons.  In
such a communication chain, the information is more likely to be
propagated to the persons that have many friends.  Random walks on a
network have similar characteristic that the probability of the random
walker arriving at a node is proportional to its node
degree~\cite{MFPT}, which corresponds to the number of friends in a
social network.  Hence, we model the information propagation on social
networks as a random walk.  However, the information propagation
modeled with the random walk may be too slower than the information
propagation in an actual social network since the random walker
arrives at the same node multiple times.  Therefore, our analysis
focuses on the worst-case situation for the information propagation in
social networks.

Random walk on network $G$ is formulated by normalized Laplacian
matrix $\bm{N}$.  When node $i$ selects node $j$ with the probability
$w_{ij}/d_i$, arrival probability $x_{a:i}(t)$ of the random walker
starting from node $a$ to node $i$ at time $t$ is given by
\begin{align}
  x_{a:i}(t) = \sum_{j \in \partial i} \frac{w_{ji}}{d_j} x_{a:j}(t-1),
\end{align}
where $\partial i = \{ j \ |\ A_{ij} > 0,\  1 \le j \le n\}$.
With arrival probability vector $\bm{x}_a(t) = (x_{a:i}(t))_{1 \le i \le n}$, we obtain
\begin{align}
  \bm{x}_a(t) &= \bm{x}_a(t-1) \bm{D}^{-1} \bm{A} \nonumber\\
  \bm{x}_a(t) \bm{D}^{-1/2} &= \bm{x}_a(t-1) \bm{D}^{-1}\bm{A}\,\bm{D}^{-1/2} \nonumber\\
  \bm{y}_a(t) &= \bm{y}_a(t-1) (\bm{I} - \bm{N}),
\end{align}
where $\bm{y}_a(t) = \bm{x}_a(t) \bm{D}^{-1/2}$.

A fundamental metric to evaluate of the information propagation speed
in a social network is {\it a first arrival time}, which is the time
required until the information is first propagated to a person.  Such
a first arrival time of the information corresponds to the time until
the random walker first arrives at a node in the random walk.  The
existing study~\cite{MFPT} derived first arrival time $f_{a:i}$ of the
random walker starting from node $a$ to node $i$ as
\begin{align}
   f_{a:i} = 2\,|E| \sum_{l=2}^{n} \frac{1}{\lambda_l} \left(\frac{q_l^2(i)}{d_i} - \frac{q_l(a)\,q_l(i)}{\sqrt{d_a\,d_i}}\right),
           \label{eq:avg_arr-time}
\end{align}
where $q_l(i)$ is the $i$-th element of eigenvector $\bm{q}_l$ of
normalized Laplacian matrix $\bm{N}$.  Since the steady-state
probability of the random walk arriving at node $i$ is given by
$d_i/(2\,|E|)$, expected value $m$ of first arrival time $f_{a:i}$ is
derived as
\begin{align}
  m &= \sum_{i=1}^n \frac{d_i}{2\,|E|} f_{a:i} \nonumber\\
  &= \sum_{i=1}^n \sum_{l=2}^{n} \frac{1}{\lambda_l} \left(q_l^2(i) - \frac{q_l(a)\,q_l(i)}{\sqrt{d_a}}\,\sqrt{d_i} \right) \nonumber\\
  &= \sum_{l=2}^{n} \frac{1}{\lambda_l} \left( \sum_{i=1}^n  q_l^2(i) - \frac{q_l(a)}{\sqrt{d_a}} \sum_{i=1}^n  q_l(i) \sqrt{d_i}\right) \nonumber\\
  &= \sum_{l=2}^{n} \frac{1}{\lambda_l} \left( \sum_{i=1}^n  q_l^2(i) - \frac{\sqrt{2\,|E|}\,q_l(a)}{\sqrt{d_a}} \sum_{i=1}^n  q_l(i)\,q_1(i) \right) \nonumber\\
  &= \sum_{l=2}^{n} \frac{1}{\lambda_l}.
  \label{eq:MFPT_origin}
\end{align}
When we derived the above equation, we used $q_1(i) =
\sqrt{d_i/(2\,|E|)}$ and the property that $\bm{q}_l$ is the
orthonormal basis.  From the above equation, we find that $m$ can be
calculated with only eigenvalue $\lambda_l$ of normalized Laplacian
matrix $\bm{N}$, and does not depend on starting node $a$.  In our
analysis, we use $m$ as the information propagation metric of the
social network.

\subsection{The Information Propagation Speed with the Universality of Normalized Laplacian Matrix $\bm{N}$}

Using the universality shown in Sect.~\ref{sec:val}, we analyze the
information propagation speed.  According to the universality,
eigenvalue frequency distribution $f_N(\lambda)$ of $\bm{N}$ for
randomly-weighted networks is approximated by $\rho_N(\lambda)$, which
given by Eq.~(\ref{eq:semicircle}).  We first derive $\tilde{m}$ that
is the approximated value of $m$ with the assumption that
$f_N(\lambda) \simeq \rho_N(\lambda)$.  Then, we discuss the
information propagation speed on the basis of $\tilde{m}$.

If $f_N(\lambda) \simeq \rho_N(\lambda)$, expected value $m$ of first
arrival time $f_{a:i}$ is approximated by
\begin{align}
  m &= \sum_{l=2}^{n} \frac{1}{\lambda_l} \simeq (n-1) \sum_{i=1}^{n_h} \frac{1}{\theta_i} f_N(\theta_i) \nonumber\\
    &\simeq (n-1)\int_{\lambda_2}^{\lambda_n} \frac{1}{\lambda} \,\rho_N(\lambda) \, \dd \lambda \nonumber\\
    &= \frac{2\,(n-1)}{\pi\,r^2} \int_{\lambda_{2}}^{\lambda_{n}} \frac{1}{\lambda} \sqrt{r^2 - (\lambda-1)^2} \,\dd \lambda \nonumber\\
    &= \frac{2\,(n-1)}{r^2} \left(1 - \sqrt{1 - r^2}\right) = \tilde{m},
    \label{eq:avg_avg_arr-time}
\end{align}
where $\theta_i = \lambda_2 + (i-1/2)\,h_b$ and $h_b = (\lambda_n -
\lambda_2)/2$.  The detailed deviation process of
Eq.~(\ref{eq:avg_avg_arr-time}) is provided in the appendix.
According to Eq.~(\ref{eq:avg_avg_arr-time}), $\tilde{m}$ is only
determined by the number $n$ of nodes, and radius $r$ of the
semicircle distribution.

On the basis of Eq.~(\ref{eq:avg_avg_arr-time}), we analyze the
worst-case speed of the information propagation in a social network.
Figure~\ref{fig:r-avg_m} shows $\tilde{m}$ given by
Eq.~(\ref{eq:avg_avg_arr-time}) as the function of radius $r$.  Note
that $r$ is within the range $(0, 1)$ because $0 < \lambda_2,
\lambda_n < 2$.  From this figure, $\tilde{m}$ is the monotonically
increasing function of $r$ because of $\dd \,\tilde{m}/\dd \, r > 0$
for $0 < r < 1$.  Hence, if $m$ is able to be approximated by
$\tilde{m}$, the worst-case speed of the information propagation in a
social network becomes slower as $r$ increases.  Moreover, since the
range of $r$ is $0 < r < 1$, the lower bound and upper bound of
$\tilde{m}$ are given by $n - 1$ and $2\,(n-1)$, respectively.  Hence,
if the structure~(i.e., relationship among people) of a social network
changes, the worst-case speed of the information propagation changes
at most 2.

\insertfig{r-avg_m}{Radius $r$ vs. $\tilde{m}$}

\subsection{The Validity of Our Analysis}

Our analysis is valid if the difference between $m$ and $\tilde{m}$ is
sufficiently small.  To investigate the difference, we use relative
error $\epsilon_m$, which is defined by
\begin{align}
\epsilon_m = \frac{|\tilde{m} - m|}{m}.
\label{error_MFPT}
\end{align}

In the investigation, we repeat the generation of $\bm{N}$ 100 times,
and calculate the average of related error $\epsilon_m$.
Like Sect.~\ref{sec:val}, we use the the parameter configuration shown
in Tab.~\ref{tab:param} as default, and randomly-weighted ER or BA
networks.  These networks have no cluster that is the set of
densely-connected nodes, and is often observed in an actual social
network.  In general, the information propagation in a cluster is very
fast, and so a cluster can be a node in the information propagation.
Hence, the investigation with BA and ER networks is also valid for
actual social networks, which may have clusters.

Figure~\ref{fig:err-m_ba_vs_er_N=1000Xk=40} shows relative error
$\epsilon_m$ for randomly-weighted ER and BA networks.  According to
this figure, $\epsilon_m$ approaches to 0 as $k_{\rm ave}$ increases
regardless of the link weight distribution and network topology.
Although the discretization error due to $n$ and $n_h$ affects
relative error $\epsilon$ of eigenvalue frequent distribution
$f_N(\lambda)$, it does not affect $\epsilon_m$.  This characteristic
is useful for the information propagation analysis.

By the comparison between
Figs.~\ref{fig:avg-deg_vs-min-deg_ba_vs_er_N=1000Xk=40} and
\ref{fig:err-m_ba_vs_er_N=1000Xk=40}, we conclude that our analysis
based on Eq.~(\ref{eq:avg_avg_arr-time}) is valid for social networks
if the degree condition $k^{2}_\mathrm{min} \gg k_\mathrm{ave}$ is
fulfilled.

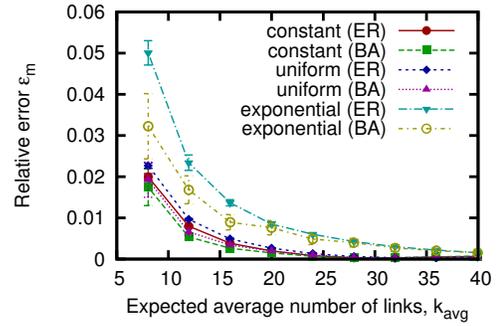
\insertfig{err-m_ba_vs_er_N=1000Xk=40}{Relative error $\epsilon_m$ of
  expected value of first-arrival time $f_{a:i}$ for randomly-weighted
  ER and BA networks}

\section{Conclusion and Future Work}
\label{sec:conclusion}

Spectral graph theory cannot be simply applied to social network
analysis because the matrix elements used in the theory cannot be
given exactly to represent the structure of a social network.  For
this reason, we first discussed the universality of random matrix with
the feature of social networks.  As such random matrix, we used
normalized Laplacian matrix $\bm{N}$ for a network where link weights
are randomly given.  We clarified that the universality~(i.e., the
Wigner semicircle law given by Eq.~(\ref{eq:semicircle})) of
normalized Laplacian matrix $\bm{N}$ appears regardless of the link
weight distribution in ER networks and BA networks.  According to the
universality, eigenvalue frequency distribution $f_N(\lambda)$ of
$\bm{N}$ is is determined by only the number $n$ of nodes and
semicircle radius $r = 1 - \lambda_2$ or $\lambda_n - 1$ where
$\lambda_2$ and $\lambda_n$ are the second minimum eigenvalue and the
maximum eigenvalue of $\bm{N}$, respectively.  Then, we analyzed the
information propagation speed in a social network on the basis of the
spectral graph theory and the clarified universality.  In this
analysis, we modeled the information propagation by a random walk in
the light of the resemblance between their characteristics, and
investigated expected value $m$ of first arrival times of the random
walker for each node.  Our analysis showed that the worst-case speed
of the information propagation changes at most 2 if the
structure~(i.e., relationship among people) of a social network
changes.

As future work, we will investigate the relationship between
topological property~(e.g., scale-free property) and radius $r$ since
$r$ determines the information propagation speed in a social network.
Then, we are planning to design a social media for effective
information propagation on the basis of the finding by our work.

\section*{Acknowledgment}
\label{sec:acknowledgements}

This work was supported by JSPS KAKENHI Grant Number 15K00431.

\renewcommand{\theequation}{A.\arabic{equation}}
\setcounter{equation}{0}

\section*{Appendix}
\addcontentsline{toc}{section}{Appendix}
\label{sec:app}

We describe the detailed deviation process of
Eq.~(\ref{eq:avg_avg_arr-time}).  If eigenvalue frequency distribution
$f_N(\lambda)$ is given by the Wigner semicircle law, expected value
$m$ of the first-arrival time is approximated by $\tilde{m}$.
$\tilde{m}$ is given by
\begin{align}
   \tilde{m} &= (n-1)\int_{\lambda_2}^{\lambda_n} \frac{1}{\lambda} \,\rho_N(\lambda) \, \dd \lambda \nonumber\\
   &= \frac{2\,(n-1)}{\pi\,r^2} \int_{\lambda_{2}}^{\lambda_{n}} \frac{1}{\lambda} \sqrt{r^2 - (\lambda-1)^2} \,\dd \lambda.
   \label{eq:adx_1}
\end{align}
By substituting $\lambda = r\,\cos \theta + 1$ into
Eq.~(\ref{eq:adx_1}), we obtain
\begin{align}
    \tilde{m} &= \frac{2\,(n-1)}{\pi\,r^2} \int_{0}^{\pi} \frac{r^2\,\sin^2\theta}{r\,\cos\theta+1} \, \dd \theta \nonumber\\
      &= \frac{2\,(n-1)}{\pi\,r^2} \int_{0}^{\pi} \frac{r^2 - 1 + 1 - r^2\,\cos^2\theta}{r\,\cos\theta+1} \, \dd \theta \nonumber\\
      &= \frac{2\,(n-1)}{\pi\,r^2} \int_{0}^{\pi} \left( \frac{r^2 - 1}{r\,\cos\theta+1} + 1 - r\,\cos\theta \right) \, \dd \theta \nonumber\\ 
      &= I_1(r) + I_2(r),
      \label{eq:adx_2}
\end{align}
where $I_1(r)$ and $I_2(r)$ are given by
\begin{align}
   I_1(r) &= \frac{2\,(n-1)}{\pi\,r^2} \int_{0}^{\pi} \frac{r^2 - 1}{r\,\cos\theta+1}\, \dd \theta, \label{eq:l1}\\
   l_2(r) &= \frac{2\,(n-1)}{\pi\,r^2} \int_{0}^{\pi} \left(1 - r\,\cos\theta \right) \, \dd \theta, \label{eq:l2}
\end{align}
respectively.  When we use the half-angle formula of $\cos\theta$,
$l_1(r)$ is given by
\begin{align}
     I_1(r) &= \frac{2\,(n-1)}{\pi\,r^2} \nonumber \\
              &\quad \times \int_{0}^{\pi} \frac{r^2 - 1}{r\left(\cos^2\frac{\theta}{2} - \sin^2\frac{\theta}{2}\right) + \left(\cos^2\frac{\theta}{2} + \sin^2\frac{\theta}{2}\right)} \, \dd \theta \nonumber\\
      &= \frac{2\,(n-1)}{\pi\,r^2} \int_{0}^{\pi} \frac{r^2 - 1}{(1+r)\,\cos^2\frac{\theta}{2} + (1-r)\,\sin^2\frac{\theta}{2}} \, \dd \theta \nonumber\\
      &= \frac{2\,(n-1)}{\pi\,r^2} \int_{0}^{\pi} \frac{r^2 - 1}{(1+r) + (1-r)\,\tan^2\frac{\theta}{2}}\frac{1}{\cos^2\frac{\theta}{2}} \, \dd \theta \nonumber\\
     &= \frac{2\,(n-1)}{\pi\,r^2} \int_{0}^{\pi} \frac{r-1}{1 - \mathrm{i}^2\,\frac{1-r}{1+r}\,\tan^2\frac{\theta}{2}}\frac{1}{\cos^2\frac{\theta}{2}} \, \dd \theta,
     \label{eq:l1_2}
\end{align}
where $\mathrm{i}$ is the imaginary unit.
By substituting $x = \sqrt{\frac{1-r}{1+r}} \tan\frac{\theta}{2}$ into Eq.~(\ref{eq:l1_2}), $l_1(r)$ is derived as
\begin{align}
   I_1(r)  &= \frac{2\,(n-1)}{\pi\,r^2} \int_{0}^{\infty} \frac{-2\,\sqrt{1-r^2}}{1 - (\mathrm{i}\,x)^2} \, \dd x  \nonumber\\
           &= \frac{2\,(n-1)}{\pi\,r^2} \nonumber\\
           &\quad \times\int_{0}^{\infty} \left(-\sqrt{1-r^2}\right) \left(\frac{1}{1 + \mathrm{i}\,x} + \frac{1}{1 - \mathrm{i}\,x}\right) \, \dd x \nonumber\\
           &= \frac{2\,(n-1)}{\pi\,r^2} \left(-\sqrt{1-r^2}\right) \nonumber \\ 
           &\quad \times \Bigl[ \frac{1}{\mathrm{i}}\,\log(1 + \mathrm{i}\,x) - \frac{1}{\mathrm{i}}\,\log(1 - \mathrm{i}\,x) \Bigr]_{0}^{\infty} \nonumber\\
           &= \frac{2\,(n-1)}{\pi\,r^2} \left(\mathrm{i}\,\sqrt{1-r^2}\right) \nonumber\\
           &\quad \times \lim_{x \rightarrow \infty} \Bigl[ \log(1 + \mathrm{i}\,x) - \log(1 - \mathrm{i}\,x)\Bigr].
           \label{eq:l1_3}
\end{align}
When we use $\log z = \log|z| + \mathrm{i}\,\mathrm{arg}(z)$, $I_1(r)$ is derived as
\begin{align}
   I_1(r)  &= \frac{2\,(n-1)}{\pi\,r^2} \left(\mathrm{i}\,\sqrt{1-r^2}\right) \lim_{x \rightarrow \infty} 2\,\mathrm{i}\,\tan^{-1} x \nonumber\\
        &= -\frac{2\,(n-1)}{r^2} \sqrt{1-r^2}.
        \label{eq:l1_4}
\end{align}

Then, $l_2(r)$ is derived as
\begin{align}
   I_2(r)  &= \frac{2\,(n-1)}{\pi\,r^2} \int_{0}^{\pi} \left(1 - r\,\cos\theta \right) \, \dd \theta \nonumber\\
        &= \frac{2\,(n-1)}{\pi\,r^2} \Bigl[\theta - r\,\sin\theta \Bigr]_0^\pi \nonumber\\
        &= \frac{2\,(n-1)}{r^2}.
        \label{eq:l2_2}
\end{align}

By substituting Eqs.~(\ref{eq:l1_4}) and (\ref{eq:l2_2}) into Eq.~(\ref{eq:adx_2}), $\tilde{m}$ is given as
\begin{align}
   \tilde{m} &= I_1(r) + I_2(r) \nonumber\\
     &= \frac{2\,(n-1)}{r^2} \left(1 - \sqrt{1-r^2}\right).
     \label{eq:MFPT_3}
\end{align}


\bibliographystyle{ieeetr}
\bibliography{bib/bib}

\end{document}